\documentclass[%
	preprint,
        aps,pra,showpacs,amsmath,amssymb]{revtex4}
\usepackage{graphicx}
\newcommand{\ket}[1]{| #1 \rangle}
\newcommand{\bra}[1]{\langle #1 |}
\newcommand{\bracket}[2]{\langle #1 | #2 \rangle}

\begin{document}

\title{Determination of the border between ``shallow'' and ``deep'' 
  tunneling regions for Herman-Kluk method
  by asymptotic approach}
\author{Atushi Tanaka}
\email{tanaka@phys.metro-u.ac.jp}
\affiliation{Department of Physics, Tokyo Metropolitan University,
  Minami-Osawa, Hachioji, Tokyo 192-0397, Japan}

\begin{abstract}
The evaluation of a tunneling tail by the Herman-Kluk method, which is a 
quasiclassical way to compute quantum dynamics, is examined by asymptotic 
analysis. In the shallower part of the tail, as well as in the classically 
allowed region, it is shown that the leading terms of semiclassical 
evaluations of quantum 
theory and the Herman-Kluk formula agree, which is known as an {\em asymptotic 
equivalence}. In the deeper part, it is shown that the 
asymptotic equivalence breaks down, due to the emergence of unusual 
``tunneling trajectory'', which is an artifact of the Herman-Kluk method.
\end{abstract}


\maketitle


Even nowadays, it is still impossible to carry out serious
numerical investigation of quantum dynamics, even
for modest (e.g., 10--100) degrees of freedom systems, 
with the present state of the art of computational technology,
unless we take drastic approximations that need to be based
on good physical insights. 
As a starting point to invent such a method, it is often employed
the semiclassical approximation (i.e. asymptotic evaluation) of 
the path integral representation of 
time evolution operator (Feynman kernel)~\cite{Schulman:1981}.
The semiclassical approximation, however, has difficulties due to 
the exponential proliferation of contributing trajectories and 
caustics (see, e.g. Ref.\ \cite{Tomsovic:PRE-1993-282}) and 
due to Stokes phenomenon that requires to remove  
non-contributing complex-valued trajectories~\cite{Adachi:AP-195-45}, 
even in few degrees of freedom systems, 
when the corresponding classical system is chaotic. 
Furthermore, in the semiclassical method, the boundary conditions
of the contributing classical trajectories both for initial and
finial times, is troublesome (known as a root-search problem~%
\cite{Miller:JCP-53-3578,Miller:JCP-56-5668}), 
in computations for realistic systems such as atoms and molecules. 
In order to avoid the root-search problem, 
initial value representations (IVRs) of Feynman kernels are proposed.
The IVRs imposes only the initial conditions on the classical trajectories.
In Ref.\ \cite{Miller:JCP-53-3578}, the earliest version of IVR is
introduced by a change of integral variables to semiclassical
Feynman kernel.
A general framework of IVR is proposed by Kay~\cite{Kay:JCP-100-4377}, 
who discussed
various integral expressions (IEs) of approximate Feynman kernel,
which are composed by classical trajectories that are emitted from
real-valued initial conditions. The important guiding principle of
Kay's theory is that the leading semiclassical expressions of both an
IE and exact quantum theory must agree. This is called
{\em the asymptotic equivalence}%
~\cite{Campolieti:JCP-98-5969,Kay:JCP-100-4377}.
Furthermore, Kay argued that several
known IVRs (IEs), including 
thawed Gaussian approximation~\cite{Heller:JCP-62-1544}, 
cellular dynamics~\cite{Heller:JCP-94-2723}, and 
the Herman-Kluk (HK) formula~\cite{Herman:CP-91-27}, 
are asymptotically equivalent with quantum theory.
Nowadays, a lot of numerical investigations of quantum dynamics,
including rather realistic systems, employ IVRs, in particular,
the Herman-Kluk method~\cite{ReviewHK}.

The limitation of IVRs, however, is not clear~\cite{HKdisscussion},
in particular, in the
descriptions of classically forbidden processes, e.g., tunneling processes,
whose conventional semiclassical treatments need to take into account
the contributions from complex-valued classical trajectories.
Numerical experiments to reproduce tunneling tails  by IVR approaches 
suggest that the ``shallow'' side of tunneling tails is tractable~%
\cite{Keshavamurthy:CPL-218-189}.
On the other hand, concerning to the ``deep'' side, the IVR approaches have
difficulties.
Kay's {\em semiclassical} analysis of $\mathcal{O}(\hbar^2)$ error term of 
Herman-Kluk method reveals that the magnitude of the error is
controlled by the {\em complex-valued} classical
trajectories~\cite{Kay:JCP-107-2313}.   
However, these works are not conclusive.
First, since the tunneling tails are exponentially small, the corresponding 
error analysis also requires to treat {\em exponentially small errors}. 
Hence, the $\mathcal{O}(\hbar^2)$ error term, which is satisfactory in 
classically allowed region, is too large. 
Second, the border between the shallow side and the deep side of the tunneling 
tail has been unknown. 
In order to clarify the limitation of IVR, the identification of the border 
is inevitable.
We remind that asymptotic (i.e., semiclassical) analysis has an ability 
to treat exponentially small quantities. This suggests some asymptotic
approach may reveal the limitation of IVRs with much better accuracy.

In this paper, we examine an evaluation of a tunneling tail, by the 
Herman-Kluk formula, with asymptotic analysis.
Here we
focus on the Feynman kernel, rather than energy spectra or correlation
functions. This facilitates to identify the origin of discrepancies. 

\begin{figure}
  \includegraphics[width=8.6cm]{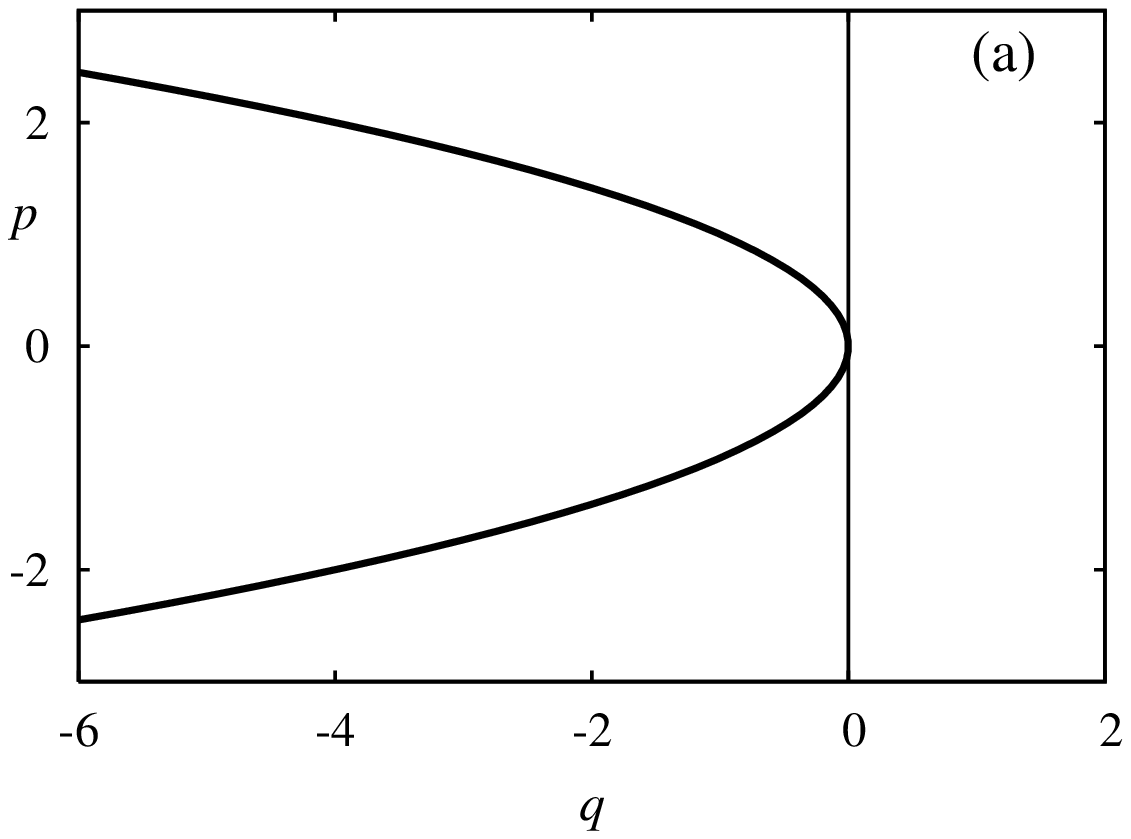}
  \hfill
  \includegraphics[width=8.6cm]{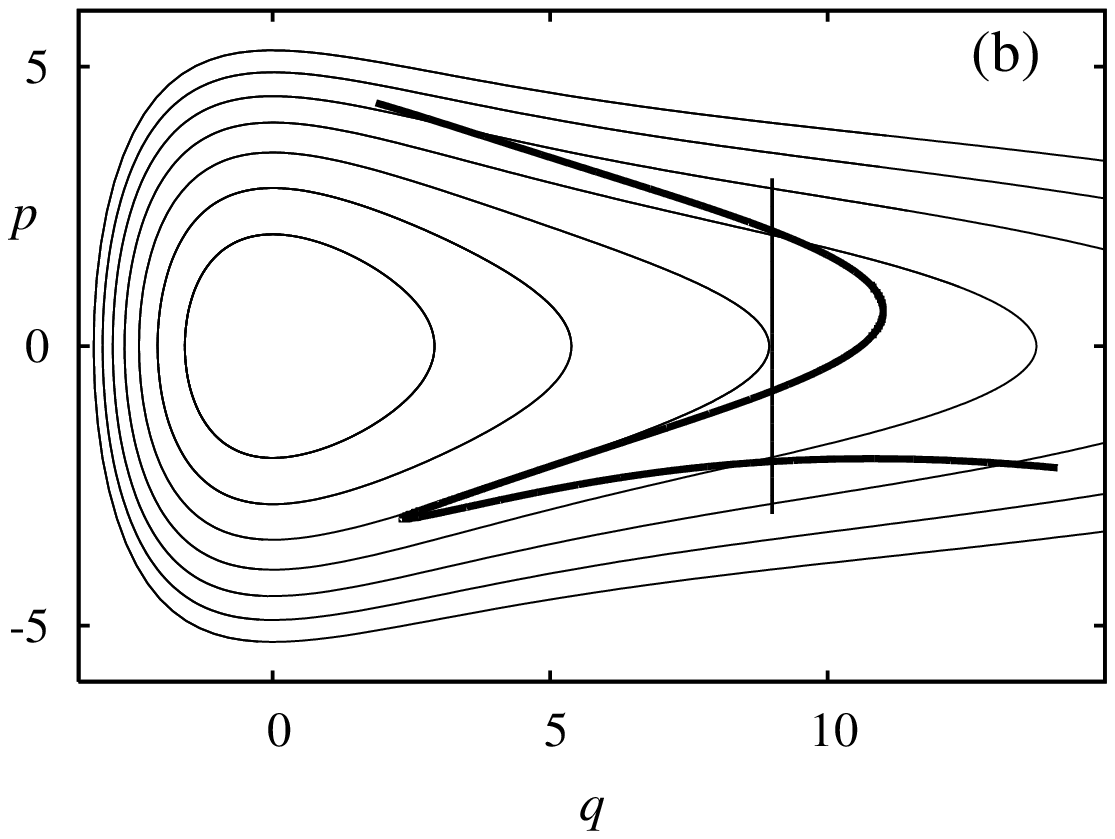}
  \caption{\label{fig:manifolds}
    Evolutions of classical manifolds in phase space.
    (a)~With the folding Hamiltonian $H = - g p^3 / 3$~\cite{Adachi:AP-195-45}.
    The initial manifold (thin line) is at $q = 0$. After a time
    interval $\tau$, the manifold folds (thick line). 
    The caustic in the position representation is at $q = 0$.
    (b)~With a nonlinear oscillator $H = \frac{1}{2} p^2 + V(q)$,
    where 
    $V(q) = D \{(1 - e^{-\lambda q})^2 - 1\} + \frac{1}{2}(1-\epsilon) q^2$,
    $\epsilon=0.975$, $\lambda = 1/\sqrt{12}$, and
    $D = \epsilon / (2 \lambda^2)$ (Contour lines of $V(q)$ are thin)%
    ~\cite{Brickmann:JCP-75-5744}. 
    The initial manifold $\{(q, p)| q = 9, -3 < p < 3\}$ (vertical line)
    mimics an eigenstate of the position operator, with an energy cutoff.
    The corresponding final manifold (thick line), at $t=18$, 
    has two prominent caustics in the position representation . 
  }
\end{figure}

We here examine a single degree of freedom system that is described by 
a Hamiltonian $H = - g p^3 / 3$, where $q$ and $p$ are the position and
the momentum of the system, respectively. We assume that the strength
of folding $g$ is positive. 
This is a canonical model that describes a
nonlinear folding process in classical phase space 
(see, Fig.\ \ref{fig:manifolds} (a))%
~\cite{Adachi:AP-195-45}.
When both the initial and the final states of Feynman kernel are
eigenstates of the position operator, the nonlinear folding dynamics
produces a caustic.
Note that in generic, nonlinear systems induce foldings in general.
Dynamics locally around each foldings are described by the canonical 
Hamiltonian with appropriate rotations and
scaling in phase space (see, e.g., Fig.~\ref{fig:manifolds}~(b)).

The Feynman kernel of a time evolution of interval $[0,\tau]$ 
($\tau > 0$) in the position representation
$K(q) \equiv \bra{q}\exp(-i H\tau/\hbar)\ket{q=0}$
is expressed exactly with Airy function 
\begin{equation}
  K(q) = A_i (q/l) / l
\end{equation}
where $l\equiv(\hbar^2 g \tau)^{1/3}$ is a characteristic length for 
a penetration into the classically forbidden region $q > 0$.
The asymptotic evaluation 
of the integral representation of the Feynman kernel
\begin{equation}
  \label{eq:exactIE}
  K(q) = \int dp \bracket{q}{p}e^{+i g p^3 \tau/ (3\hbar)}\bracket{p}{q=0}
\end{equation}
gives a leading semiclassical approximation $K_{\rm SC}$, which are 
composed by classical trajectories.
On one hand, in the classically allowed region $q < 0$, we have
a superposition of incoming and outgoing waves:
\begin{equation}
  K_{\rm SC}(q) = 
  \frac{1}{\sqrt{\pi} l (|q|/l)^{1/4}}
  \cos\left\{\frac{2}{3}\left(\frac{|q|}{l}\right)^{3/2}-\frac{\pi}{4}
  \right\}.
\end{equation}
The corresponding classical trajectories are characterized by their
momentum, which are conserved quantities: 
$p_{\pm}(q) = \pm\sqrt{|q|/(\tau g)}$.
On the other hand, in the classically forbidden region $q > 0$, 
we have a tunneling tail:
\begin{equation}
  \label{eq:KSCtail}
  K_{\rm SC}(q) = 
  \frac{1}{2\sqrt{\pi} l (q/l)^{1/4}}
  \exp\left\{-\frac{2}{3}\left(\frac{q}{l}\right)^{3/2}\right\}.
\end{equation}
The momentum of the tunneling trajectory is pure imaginary: 
$p_0(q) = i \sqrt{q/(\tau g)}$. 
Between these regions, at $q=0$, these 
classical trajectories merge to produce a caustic.
The change of asymptotic expansions for different signs of $q$ 
is controlled by Stokes phenomenon~\cite{Berry:RPP-35-315}.

\begin{figure}
  \includegraphics[width=0.3\textwidth]{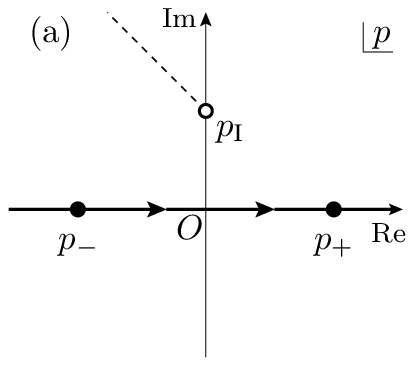}
  \hfill
  \includegraphics[width=0.3\textwidth]{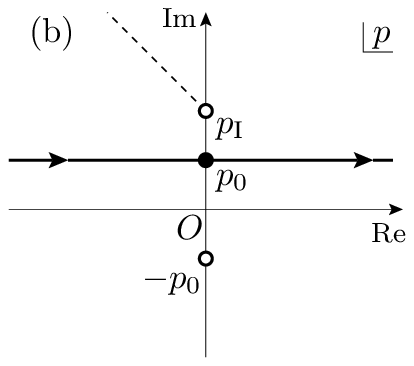}
  \hfill
  \includegraphics[width=0.3\textwidth]{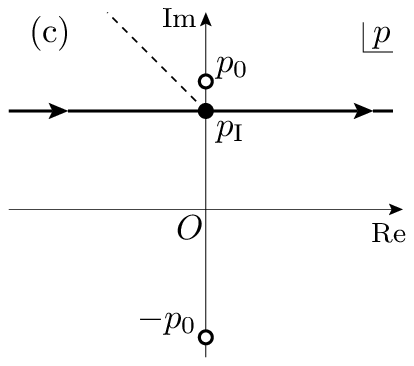}
  \caption{\label{fig:integrationPath}
    Locations of saddle points (closed and open circles) and integration paths 
    (thick lines) for the asymptotic evaluations of 
    the integral~(\ref{eq:HKp}), 
    for (a)~the classical region $q<0$, 
    (b)~the shallow tunneling region $0<q<l_{\gamma}$, and,
    (c)~the deep tunneling region  $q>l_{\gamma}$.
    Dashed lines emanating from the branch point $p_{\rm I}$ are
    branch cuts. 
    Closed and open circles are contributing and non-contributing 
    saddle points, respectively.
  }
\end{figure}

In the following, the corresponding Herman-Kluk kernel~%
\cite{Herman:CP-91-27} is examined:
\begin{eqnarray}
  \label{eq:HK}
  K^{\rm HK}(q) &=& 
  \int \int \frac{dq_0  dp_0}{2\pi\hbar}
  \bracket{q}{\varphi^{\gamma}(q_{\tau}, p_{\tau})} C(q_0, p_0, \tau) 
  \nonumber \\ 
  &&{}\times
  e^{i S_{\tau}(q_0, p_0)/\hbar}
  \bracket{\varphi^{\gamma}(q_0, p_0)}{q = 0},
\end{eqnarray}
where $\gamma$ ($>0$) determines the width of Gaussian packet
$\bracket{q}{\varphi^{\gamma}(q_0, p_0)}
= (2\gamma/\pi)^{1/4}\exp\{-\gamma (q - q_0)^2 + i p_0 (q - q_0)/\hbar\}$,
$(q_{\tau}, p_{\tau})$ is the classical trajectory at time
$t = \tau$, emitted from $(q_0, p_0)$ at $t =0$, 
$S_{\tau}(q_0, p_0)$ is the classical action along the time evolution,
and 
$C(q_0, p_0, \tau) = 
\{\partial q_{\tau}/\partial q_0 + \partial p_{\tau}/\partial p_0 
- 2i\hbar\gamma \partial q_{\tau}/\partial p_0
- (2i\hbar\gamma)^{-1} \partial p_{\tau}/\partial q_0\}^{1/2}/\sqrt{2}$.
For the folding Hamiltonian, $C(q_0, p_0, \tau)$ have a branch point
at $p_{\rm I} = i / (2\hbar\gamma\tau g)$. After the integration of 
the variable
$q_0$, $K^{\rm HK}(q)$~(\ref{eq:HK}) becomes
\begin{equation}
  \label{eq:HKp}
  K^{\rm HK}(q)
  = \int \frac{dp}{2\pi\hbar} C(p, \tau) e^{-\phi_{\tau}(p)},
\end{equation}
where $C(p,\tau) = (1 - p/p_{\rm I})^{1/2}$ and
$\phi_{\tau}(p) = 
\gamma (q + \tau g p^2)^2/2 - i p q /\hbar -i \tau g p^3 / (3\hbar)$.
The integral~(\ref{eq:HKp}) has three saddle points
$p = \pm\sqrt{- q/(\tau g)}$ and $p_{\rm I}$. The former momenta
$p = \pm\sqrt{- q/(\tau g)}$ correspond to
the classical momenta $p_{\pm}(q)$ in the classically allowed region and
$p_0(q)$ in the classically forbidden region. 
The latter momentum  $p_{\rm I}$, which is
the branch 
point of $C(p,\tau)$, appears only in the semiclassical 
analysis of $K^{\rm HK}(q)$. 
Note that all the saddle points need not to make contributions
to the semiclassical kernel, due to the Stokes phenomena.
Actually, 
in the classically allowed region $q>0$, there are the contributions
only from  $p = p_{\pm}(q)$ (FIG.~\ref{fig:integrationPath}(a)). 
Hence the leading semiclassical
evaluation of $K^{\rm HK}(q)$ agree with $K_{\rm SC}(q)$. Thus the
asymptotic equivalence between Herman-Kluk kernel
and quantum theory holds for $q<0$~\cite{Kay:JCP-100-4377}.

In the semiclassical evaluation of $K^{\rm HK}(q)$ in the classically
forbidden region $q>0$, there is a length scale 
$l_{\gamma} = \gamma^{-2} l^{-3} / 4$ that divides the tunneling tail
into two regions: a ``shallow'' region $0 < q < l_{\gamma}$ and 
a ``deep'' region $q > l_{\gamma}$.
In the shallow region, the semiclassical evaluation of $K^{\rm HK}(q)$
has only a single contribution from the classical trajectory 
$p = p_0(q)$ (FIG.~\ref{fig:integrationPath}(b)). 
Hence the asymptotic equivalence between 
Herman-Kluk kernel and quantum theory holds both classically allowed
region and the shallow tunneling region $q < l_{\gamma}$. 
This is a promising evidence that collections of real classical
trajectories can describe {\em shallow} tunneling tails, through
IVR techniques.
This is the first example to show that IVR technique can
describe a classically forbidden process with an analytical argument.

On the other hand, in the ``deep'' tunneling region, we found a
discrepancy: Due to Stokes phenomenon, the contribution from
``conventional'' tunneling trajectory $p = p_0(q)$ disappears, and in
turn, the contribution from the classical trajectory $p = p_{\rm I}$,
which is an artifact of the Herman-Kluk kernel, appears
(FIG.~\ref{fig:integrationPath}(c)).
The resultant semiclassical evaluation of $K^{\rm HK}(q)$ is
\begin{eqnarray}
  \label{eq:HKSCtail}
  K^{\rm HK}_{\rm SC}(q) &=& 
  \frac{\Gamma(3/4)}{2\pi l (\gamma l^2)^{1/4}}
  \left(\frac{l}{q - l_{\gamma}}\right)^{3/4}
  \nonumber\\ 
  &&{}\times
  \exp\left\{-\frac{\gamma}{2}(q + l_{\gamma})^2 -
    \frac{4}{3}l_{\gamma}^2\right\}.
\end{eqnarray}
Note that $q = l_{\gamma}$ is a caustic, which is an reminiscent of
the Stokes phenomena. At the same time, the asymptotic form of
tunneling tail $\sim \exp(-\gamma q^2 /2)$ for $q \gg l_{\gamma}$,
which sensitively depends on $\gamma$, is qualitatively different from 
$K_{\rm SC}(q)\sim\exp\{-2 (q/l)^{3/2}/3\}$~(\ref{eq:KSCtail}). 
Thus the breakdown of the asymptotic equivalence is evident.

\begin{figure}
  \includegraphics[width=8.6cm]{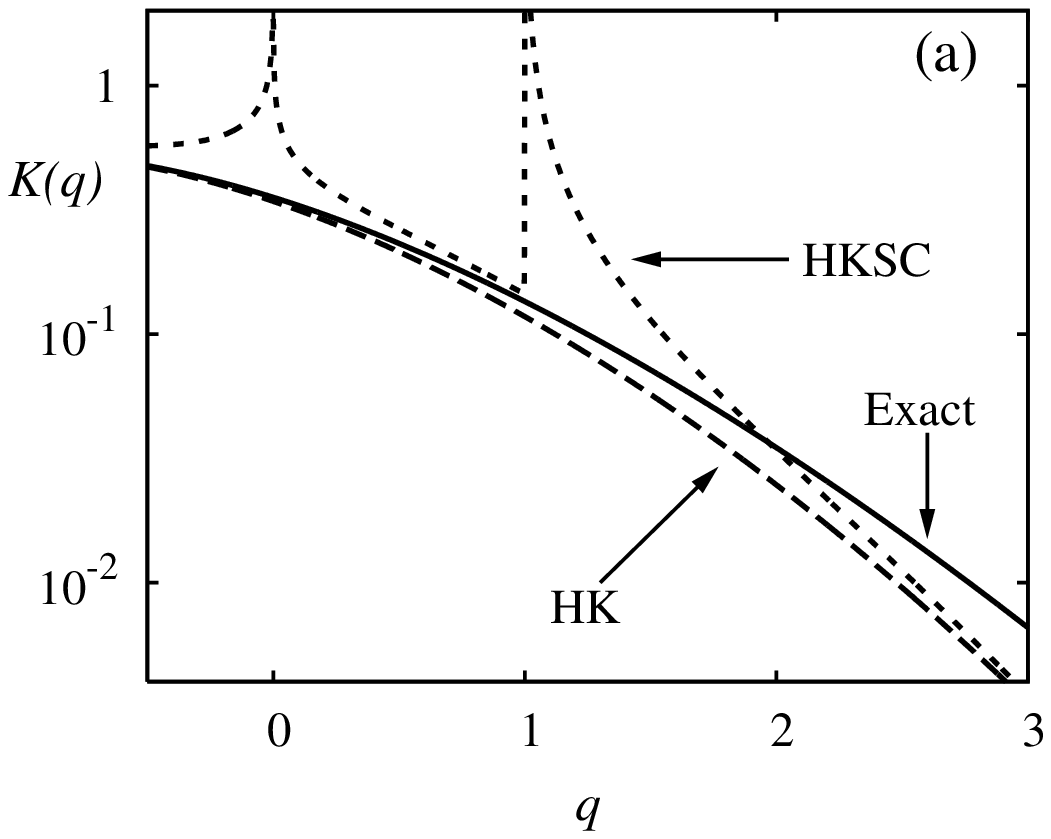}
  \hfill
  \includegraphics[width=8.6cm]{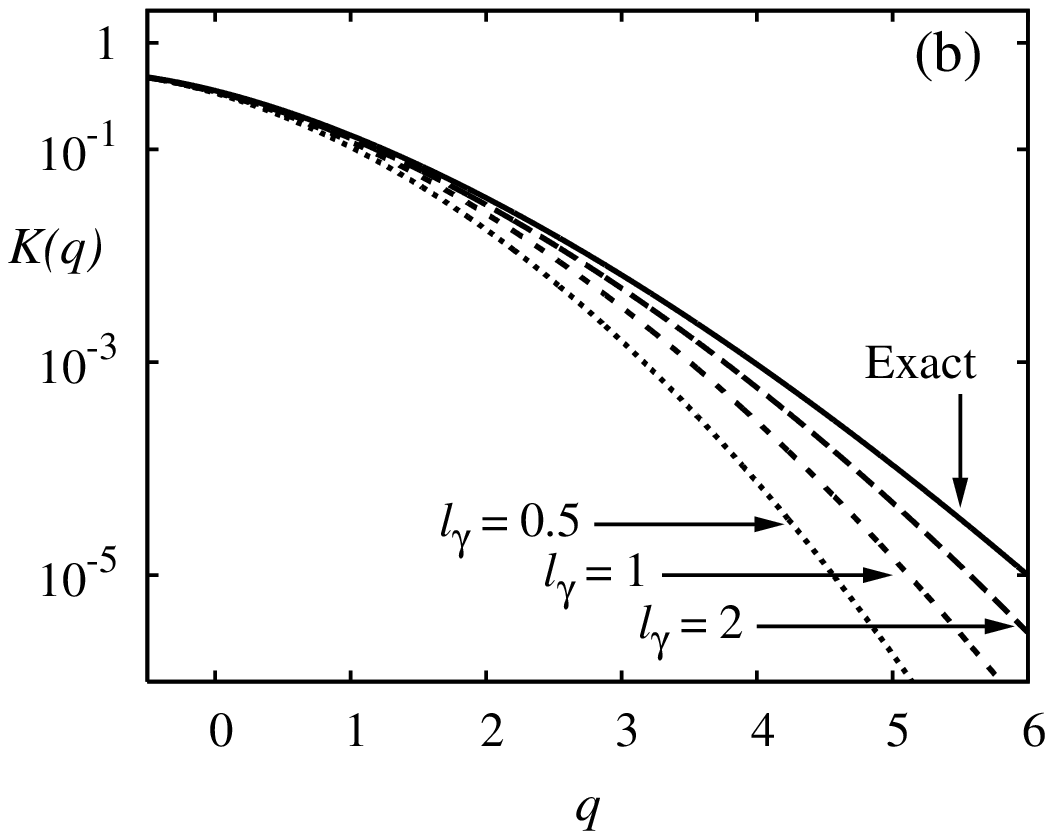}
  \caption{\label{fig:compareK}
    Comparisons of exact, Herman-Kluk (HK), and semiclassical
    Herman-Kluk (HKSC) evaluations of the Feynman kernel $K(q)$, 
    in particular,
    its tunneling tail $q > 0$. The penetration depth is $l = 1$.
    (a) With $l_{\gamma} = 1$, all three theories are shown.
    At the (conventional) turning point $q=0$, the semiclassical 
    Herman-Kluk encounters caustic.
    In the shallow region $0 < q < l_{\gamma}$, the
    discrepancy between quantum theory and Herman-Kluk method is not
    significant. On the other hand, at $q=l_{\gamma}$, the semiclassical
    Herman-Kluk encounters another caustic. Note that the conventional 
    semiclassical theory meets the caustic at $q=0$, the conventional
    turning point, only.
    In the deep region $q > l_{\gamma}$, 
    the tunneling tails of quantum theory and Herman-Kluk 
    formula take qualitatively different shape
    (see, Eq.\ (\ref{eq:KSCtail}) and Eq.\ (\ref{eq:HKSCtail})).
    (b) Comparison of quantum theory and Herman-Kluk formula, with several 
    values of $l_{\gamma}$. 
    In order to reproduce the shape of deep tunneling tail
    by Herman-Kluk method, we need larger $l_{\gamma}$~\cite{Kay:JCP-107-2313}.
  }
\end{figure}

In the argument above, the discrepancy between Herman-Kluk kernel and 
Feynman kernel comes from (1) a nonlinear folding dynamics in the
corresponding classical phase space, and (2) the appearance of
artificial tunneling trajectory. When the folding dynamics is not
significant (this is the case before Ehrenfest time), there is a
workaround to remove the contribution from the artificial tunneling 
trajectories, by 
adjusting the value of $\gamma$ in the Herman-Kluk kernel
to push $p_{\rm I}$ deeper in the complex plane
(see., Fig.\ \ref{fig:integrationPath}).
Indeed, this is a known strategy, which is proposed by Kay,
to reduce the magnitude of the error 
of Herman-Kluk method, and the strategy succeeds to 
a certain extent~\cite{Kay:JCP-107-2313,Maitra:JPC-112-531}.
However, in generic, nonlinear systems, it will become
difficult to carry out such workarounds in practice, due to
the emergence of multiple foldings 
(see, e.g., Fig.\ \ref{fig:manifolds}~(b)).

We summarize this paper.
With an exactly solvable
model that describes nonlinear folding process in corresponding
classical phase-space dynamics, we identified a boundary between a
shallow and deep tunneling regions for the Herman-Kluk kernel.
In the former region, the Herman-Kluk kernel and quantum theory are
asymptotically equivalent. Hence there remains a hope that Herman-Kluk
kernel describes classically forbidden process to a certain extent.
However, in the deep region, the breakdown of the asymptotic
equivalent is shown. 
Besides Herman-Kluk method, other IVR approaches, in particular, which
are based on Kay's framework~\cite{Kay:JCP-100-4377}, will have similar 
scenario on successes and failures in descriptions of classically 
forbidden phenomena.
We expect that the nonlinear folding Hamiltonian employed
here is a good test-case not only for the Herman-Kluk method, but
also for various semiclassical method. We remind that this model was 
employed to elucidate the limitation of single trajectory approximation of
semiclassical coherent-state path integrals~\cite{Adachi:AP-195-45,%
Klauder:PRL-1986-897}.
At the same time, it is highly desirable to develop a convenient way to
find the boundary between the shallow and the deep regions of
Herman-Kluk method, though the present analysis requires a
full knowledge of the Stokes phenomena to determine the boundary.

\begin{acknowledgments}
This work is partially supported by the Grant-in-Aid for Young Scientists (B) 
(No.~15740241) from the Ministry of Education, Culture, Sports, Science 
and Technology, Japan.
\end{acknowledgments}




\end{document}